\documentclass[citeauthoryear]{llncs}
\usepackage{hyperref}
\usepackage{helvet}
\usepackage{amssymb}
\usepackage{amsmath}
\usepackage{helvet}
\usepackage{algorithm}
\usepackage[noend]{algpseudocode}
\usepackage{graphicx}
\usepackage{epstopdf}
\usepackage{subcaption}
\usepackage{array}
\usepackage{float}
\usepackage{rotating}
\usepackage{hyperref}
\usepackage{hhline}
\usepackage{color, colortbl}

\usepackage{tikz}

\floatstyle{plaintop}
\restylefloat{table}
\newcolumntype{R}[1]{>{\centering\raggedleft\let\newline\\\arraybackslash\hspace{0pt}}m{#1}}

\begin{document}

\title{A branch and bound algorithm for a fractional 0-1 programming problem}

\author{Irina Utkina \and Mikhail Batsyn\inst{*} \and Ekaterina Batsyna}
\institute{
Department of Applied Mathematics and Informatics, \\Laboratory of Algorithms and Technologies for Network Analysis\\
National Research University Higher School of Economics\\
136 Rodionova street, Niznhy Novgorod, Russia\\ 
\email{iutkina@hse.ru, mbatsyn@hse.ru, batcyna@hse.ru}
}

\maketitle

\begin{abstract}
We consider a fractional 0-1 programming problem arising in manufacturing.
The problem consists in clustering of machines together with parts processed on these machines into manufacturing cells so that intra-cell processing of parts is maximized and inter-cell movement is minimized.
This problem is called Cell Formation Problem (CFP) and it is an NP-hard optimization problem with Boolean variables and constraints and with a fractional objective function. Because of its high computational complexity there are a lot of heuristics developed for it. In this paper we suggest a branch and bound algorithm which provides exact solutions for the CFP with a variable number of cells and grouping efficacy objective function. This algorithm finds optimal solutions for 21 of the 35 popular benchmark instances from literature and for the remaining 14 instances it finds good solutions close to the best known.
\end{abstract}

\begin{keywords}
cell formation; biclustering; branch and bound; upper bound; exact solution
\end{keywords}

\section{Introduction}
The first work on the Group Technology in manufacturing was written by Flanders (\cite{Flanders}). In Russia the Group Technology was introduced by Mitrofanov (\cite{Mitrofanov}). The main problem in the Group Technology (GT) is to find an optimal partitioning of machines and parts into manufacturing cells, in order to maximize intra-cell processing and minimize inter-cell movement of parts. Maximization of the so-called grouping efficacy is accepted in literature as a good objective combining these two goals (Kumar \& Chandrasekharan, \cite{3}). This problem is called the Cell Formation Problem (CFP) (Goldengorin et al., \cite{Goldengorin}). CFP with grouping efficacy objective function is a fractional 0-1 programming problem.

Burbidge developed Product Flow Analysis (PFA) approach to this problem and described the GT and the CFP in his book (Burbidge, \cite{Burbidge_1}).
Ballakur \& Steudel (\cite{Ballakur_Steudel}) have shown that the CFP is an NP-hard problem for different objective functions. That is why there have been developed a lot of heuristic approaches (Goncalves \& Resende, \cite{19}; James et al., \cite{4}; Bychkov et al., \cite{Bychkov_heur}, Paydar \& Saidi-Mehrabad, \cite{2}) and almost no exact ones for the CFP with a variable number of cells and grouping efficacy objective function.

Kusiak et al. (\cite{22}) consider one of the most simple variants of the CFP called the machine partitioning problem in which it is necessary to partition only machines into the specified number of cells minimizing the total Hamming distance between machines inside the cells. The authors present an exact A* algorithm for this variant of the CFP. They also develop a branch and bound algorithm for the CFP with a variable number of cells, a limit on the number of machines inside each cell, and maximization of the size of so-called mutually separable cells as an objective function. Spiliopoulos \& Sofianopoulou (\cite{23}) and Arket et al. (\cite{21}) also present branch and bound algorithms for the machine partitioning problem.

One of the recent exact approaches for the CFP with the grouping efficacy objective function is suggested by Elbenani \& Ferland (\cite{1}). These authors suggest to reduce the fractional programming CFP problem to a number of ILP problems by means of Dinkelbach approach and to solve each ILP problem with CPLEX solver. Unfortunately they consider the CFP with a fixed number of cells which is much easier. They solve 27 of the 35 popular benchmark instances, but only for a fixed number of cells. The same simplified formulation of the CFP is considered by Brusco (\cite{51}). The author develops a branch and bound algorithm and solves 31 of the 35 instances, but again only for some fixed numbers of cells. For example problem 26 is solved only for 7 cells and it requires more than 15 days of computational time.

To the best of our knowledge the only existing exact approach to the CFP with a variable number of cells and grouping efficacy objective function is by Bychkov et al. (\cite{Bychkov}) who suggested a new approach to reduce the CFP problem to a small number of ILP problems and for the first time solved to optimality 14 of the 35 popular benchmark instances from literature using CPLEX software.
Zilinskas et al. (\cite{Zilinskas}) considered the CFP with a variable number of cells as a bi-objective optimization problem and developed an exact algorithm which finds Pareto frontier.

In this paper we suggest an efficient branch and bound algorithm for the CFP with a variable number of cells and grouping efficacy objective function.
We are able to find optimal solutions for 21 of the 35 benchmark instances.
Note also that the CFP is a biclustering problem in which we simultaneously cluster machines and parts into cells.
So the suggested approach can be also applied to biclustering problems arising in data mining (Busygin et al., \cite{Busygin}).

\section{Formulation}
The objective of the CFP is to find an optimal partitioning of machines and parts into groups (production cells, or shops) in order to minimize the inter-cell movement of parts from one cell to another and to maximize intra-cell processing operations. The input data for this problem is matrix $A$ which contains zeroes and ones. The size of this matrix is $m \times p$ which means that it has $m$ machines and $p$ parts. The element $a_{ij}$ of the input matrix is equal to one if part $j$ should be processed on machine $i$.  The objective is to minimize the number of zeroes inside cells and the number of ones outside cells. There have been suggested several objective functions which combine these two goals.
The objective function which provides a good combination of these goals and is widely accepted in literature is the grouping efficacy suggested by Kumar \& Chandrasekharan (\cite{3}):
\begin{equation}
f = \frac{n_1^{in}}{n_1+n_0^{in}} \to \max,
\end{equation}
where $n_1$ is the number of ones in the input matrix, $n_1^{in}$ is the number of ones inside cells, $n_0^{in}$ is the number of zeroes inside cells.

The mathematical programming model for the CFP is the following (see also Bychkov et al. (\cite{Bychkov})).\\
Decision variables:
\begin{equation}
x_{ik} =
    \begin{cases}
    1 & \text{if machine } i \text{ is assigned to cell } k\\
    0 & \text{otherwise}\\
    \end{cases}
\end{equation}
\begin{equation}
y_{jk} =
    \begin{cases}
    1 & \text{if part } j \text{ is assigned to cell } k\\
    0 & \text{otherwise}\\
    \end{cases}
\end{equation}
Objective function:
\begin{equation}
    \max \frac{n_1^{in}}{n_1+n_0^{in}}
\end{equation}
Constraints:
\begin{equation}
n_1^{in} = \sum_{k=1}^{c} \sum_{i=1}^{m} \sum_{j=1}^{p} a_{ij}x_{ik}y_{jk}
\end{equation}
\begin{equation}
n_0^{in} = \sum_{k=1}^{c} \sum_{i=1}^{m} \sum_{j=1}^{p} (1-a_{ij})x_{ik}y_{jk}
\end{equation}
\begin{equation}
\label{re_x}
\sum_{k=1}^{c} x_{ik} = 1 \quad \forall i = 1,\ldots,m
\end{equation}
\begin{equation}
\label{re_y}
\sum_{k=1}^{c} y_{jk} = 1 \quad \forall j = 1,\ldots,p
\end{equation}
\begin{equation}
\label{re_x_2}
\sum_{i=1}^{m} \sum_{j=1}^{p} x_{ik}y_{jk} \ge \sum_{i=1}^{m} x_{ik} \quad \forall k = 1,\ldots,c
\end{equation}
\begin{equation}
\label{re_y_2}
\sum_{i=1}^{m} \sum_{j=1}^{p} x_{ik}y_{jk} \ge \sum_{j=1}^{p} y_{jk} \quad \forall k = 1,\ldots,c
\end{equation}
Here $c = \min(m,p)$ is the maximum possible number of cells. Constrains \eqref{re_x} and \eqref{re_y} require that every machine and every part is assigned to exactly one cell. Constrains \eqref{re_x_2} and \eqref{re_y_2} require that there are no cells having only machines without parts or only parts without machines.

\section{Branch and bound algorithm}
\subsection{Branching}

Because of the biclustering structure of the CFP our branching goes by two parameters. The suggested algorithm has branching on machines and parts sequentially changing each other: machines-parts-machines-...
We use vectors $M (1 \times m)$ and $P (1 \times p)$ for this purpose.
Element $M_i$ contains the cell to which machine $i$ is assigned and element $P_j$ contains the cell to which part $j$ is assigned.
For example M = [1231] and P = [11321] mean that cell 1 contains machines 1, 4 and parts 1, 2, 5, cell 2 contains machine 2 and part 4, and cell 3 contains machine 3 and part 3.

Branching on machines makes changes in vector $M$. It starts from assigning the first machine to cell 1.
Let $k$ be the number of cells in the current partial solution.
When the algorithm branches on machines, it takes the first machine which is not assigned to any cell and tries to assign it to the existing cells with numbers from $1$ to $k$ or creates a new cell $(k+1)$ for this machine.

Branching on parts makes changes in vector $P$. It starts with all zeroes inside $P$ which means that no parts are assigned to any cell. When the algorithm branches on parts it takes the first part which is not assigned to any cell and tries to assign it to the existing cells from $1$ to $k$ or to a new cell $(k+1)*$ (star means that the number of the cell can be $k+1$ or greater) if there are some unassigned machines which can be also added later to this new cell. We assume that the number of parts is greater than the number of machines.

The algorithm branches on parts and machines successively. It starts with $M = [100\dots0]$ and $P=[00\dots0]$. Next it changes vector $P$, then - vector $M$ and so on. This way the algorithm builds the search tree. The leaves of the search tree contain complete solutions and other nodes contain partial solutions.
The complete search tree depends only on the number of machines and parts. It contains all feasible solutions as its leaves.

To provide an efficient branching, before choosing a branch we calculate an upper bound for each branch and choose the branch with the greatest value of the upper bound. This branching strategy allows us to find good solutions earlier.

\subsection{Upper bound}
To obtain an upper bound for a given partial solution we relax the original CFP problem and suggest a polynomial algorithm to calculate an optimal solution or an upper bound for the relaxed problem.
The relaxed problem is formulated as follows.
We are given a partial solution in which some of the machines and parts are already assigned to some cells.
For example in Table \ref{partial} machines 1, 2 with parts 1, 2, 3, 4 are assigned to cell 1, and machine 3 with part 5 is assigned to cell 2.
The objective is to assign the remaining machines independently on each other to the existing cells or to a new cell, and assign the remaining parts to the existing cells taking into account only the rows already assigned in the given partial solution.
In the relaxed problem we allow an independent assignment of machines and parts to cells.
In this case the best assignment for machine 4 will be to put it to cell 1 with parts 1, 2, 3, 4, 7.
This will bring 4 ones and 1 zero inside cells.
The best assignment for machine 5 will be to put it to a new cell 3 with parts 7, 8.
This will bring 2 ones and 0 zeroes inside cells.
The best assignment for parts 6 and 7 which takes into account only rows 1, 2, 3 will be to put it to cell 2 (with machine 3).
The best assignment for part 8 which takes into account only rows 1, 2, 3 will be to put it to cell 1 (with machines 1, 2).
This optimal solution for the relaxed problem is shown in Table \ref{relaxed}.
This solution is infeasible for the original CFP problem because independent assignment of machines and parts is allowed and as a result we obtain non-rectangular cells which can also intersect by columns.
Since it is an optimal solution to the relaxed problem it provides an upper bound to the original problem.
In our example for the partial solution we have $f = \frac{8}{21 + 1} \approx 0.36$ and the solution of the relaxed problem gives us an upper bound to the complete solution of the CFP equal to $UB = \frac{8 + 10}{21 + 1} \approx 0.82$.

\begin{table}
\centering
        \begin{tabular}{ c|cccccccc|}
                  \multicolumn{1}{c}{~} & \multicolumn{1}{c}{1} & \multicolumn{1}{c}{2} & \multicolumn{1}{c}{3} & \multicolumn{1}{c}{4} & \multicolumn{1}{c}{5} & \multicolumn{1}{c}{6} & \multicolumn{1}{c}{7} & \multicolumn{1}{c}{8} \\
                   \hhline{~--------}
                  1 & \cellcolor{yellow}1  & \cellcolor{yellow}1 &\cellcolor{yellow}1 &\multicolumn{1}{c|}{\cellcolor{yellow}1} &\cellcolor{gray}1  &0 &0 &1\\
                  2 &\cellcolor{yellow}1 &\cellcolor{yellow}1 &\cellcolor{yellow}0 &\multicolumn{1}{c|}{\cellcolor{yellow}1} &\cellcolor{gray}0 &0 &0 &1\\
                   \hhline{~-----}
                  3 &\cellcolor{gray}0 &\cellcolor{gray}0 &\cellcolor{gray}1 &\cellcolor{gray}0 &\multicolumn{1}{|c|}{\cellcolor{green}1} &1 &1 &0\\
                   \hhline{~~~~~-}
                  4 &1 &0 &1 &1 &1 &0 &1 &0\\
                  5 &0 &0 &0 &0 &0 &0 &1 &1\\
                  \hhline{~--------}
        \end{tabular}
        \caption{A partial solution for the CFP}
\label{partial}
\end{table}

\begin{table}
\centering
        \begin{tabular}{ c|cccccccc|}
                  \multicolumn{1}{c}{~} & \multicolumn{1}{c}{1} & \multicolumn{1}{c}{2} & \multicolumn{1}{c}{3} & \multicolumn{1}{c}{4} & \multicolumn{1}{c}{5} & \multicolumn{1}{c}{6} & \multicolumn{1}{c}{7} & \multicolumn{1}{c}{8} \\
                   \hhline{~--------}
                  1 &\cellcolor{yellow}1  & \cellcolor{yellow}1 &\cellcolor{yellow}1 &\multicolumn{1}{c|}{\cellcolor{yellow}1} &1  &0 &0 &\cellcolor{yellow}1\\
                  2 &\cellcolor{yellow}1 &\cellcolor{yellow}1 &\cellcolor{yellow}0 &\multicolumn{1}{c|}{\cellcolor{yellow}1} &0 &0 &0 &\cellcolor{yellow}1\\
                   \hhline{~-----}
                  3 &0 &0 &1 &0 &\multicolumn{1}{|c|}{\cellcolor{green}1} &\cellcolor{green}1 &\cellcolor{green}1 &0\\
                  4 &\cellcolor{yellow}1 &\cellcolor{yellow}0 &\cellcolor{yellow}1 &\cellcolor{yellow}1 &1 &0 &\cellcolor{yellow}1 &0\\
                  5 &0 &0 &0 &0 &0 &0 &\cellcolor{red}1 &\cellcolor{red}1\\
                  \hhline{~--------}
        \end{tabular}
        \caption{Optimal solution for the relaxed CFP}
\label{relaxed}
\end{table}

In our example from Table \ref{partial} it is not obvious whether the chosen alternative $(a_1, b_1) = (4, 1)$ (putting 4 ones and 1 zero inside cells) is better than alternative $(a_2, b_2) = (1, 0)$ for machine 3. To choose between two alternatives we use \Call{CompareAlternatives}{} function (see Algorithm \ref{alg}).
It returns the index (1 or 2) of the best alternative among two ones, $-1$ if it cannot choose between these alternatives, or 0 if these two alternatives are equivalent.
We do not provide the proof of its correctness in this short paper.

\begin{algorithm}
\caption{Algorithm to choose between two alternatives}
\label{alg}
\begin{algorithmic}
\Function {CompareAlternatives}{$a_1, b_1, a_2, b_2, n_1, n_0, n_1^{in}, n_0^{in}, \bar{n}_1^{out}, \bar{n}_0^{out}, n_1^i, n_0^i$}
 \State $\Delta a \gets a_2 - a_1, \Delta b \gets b_2 - b_1$ \Comment{$\Delta b$ should be non-negative}

 \If {$(\Delta b = 0)$}
  \If {$(\Delta a < 0)$}
   \State \Return 1
  \ElsIf {$(\Delta a > 0)$}
   \State \Return 2
  \Else
   \State \Return 0
  \EndIf
 \EndIf

 \State $a_c \gets n_1^{in}, b_c \gets n_1 + n_0^{in}, b_l \gets b_c, b_u \gets n_1 + n_0 - \bar{n}_0^{out} - n_0^i, l \gets \frac{a_c}{b_c}, u \gets \frac{n_1 - \bar{n}_1^{out} - n_1^i}{b_c}$

 \If {$b_l \left(l - \frac{\Delta a}{\Delta b}\right) \ge b_1 \frac{\Delta a}{\Delta b} - a_1$}
  \State \Return 1
 \EndIf

 \If {$b_u \left( u - \frac{\Delta a}{\Delta b} \right) \le b_1 \frac{\Delta a}{\Delta b} - a_1$}
  \State \Return 2
 \EndIf

 \State \Return -1

\EndFunction
\end{algorithmic}
\end{algorithm}

Below we present a polynomial algorithm of calculating the suggested upper bound as an optimal solution of the relaxed CFP problem, if we can always choose the best alternative for every machine and part, or as an upper bound to this solution otherwise.
We illustrate the algorithm on the instance shown in Table \ref{ex}.

\begin{table}
\centering
        \begin{tabular}{ c|ccccccccc|}
                  \multicolumn{1}{c}{~} & \multicolumn{1}{c}{1} & \multicolumn{1}{c}{2} & \multicolumn{1}{c}{3} & \multicolumn{1}{c}{4} & \multicolumn{1}{c}{5} & \multicolumn{1}{c}{6} & \multicolumn{1}{c}{7} & \multicolumn{1}{c}{8} & \multicolumn{1}{c}{9} \\
                   \hhline{~---------}
                  1 & \cellcolor{yellow}1  & \cellcolor{yellow}1 &\cellcolor{yellow}1 &\cellcolor{yellow}1 &\multicolumn{1}{c|}{\cellcolor{yellow}1} &\cellcolor{gray}0 &\cellcolor{gray}0 &\cellcolor{blue}\color{white}0 &\cellcolor{blue}\color{white}0\\
                  2 &\cellcolor{yellow}1 &\cellcolor{yellow}1 &\cellcolor{yellow}1 &\cellcolor{yellow}1 &\multicolumn{0}{c|}{\cellcolor{yellow}0} &\cellcolor{gray}0 &\cellcolor{gray}0 &\cellcolor{blue}\color{white}0 &\cellcolor{blue}\color{white}1\\
                   \hhline{~------->{\arrayrulecolor{blue}}-->{\arrayrulecolor{black}}}
                  3 &\cellcolor{gray}0 &\cellcolor{gray}0 &\cellcolor{gray}0 &\cellcolor{gray}0 &\cellcolor{gray}0 &\multicolumn{1}{|c}{\cellcolor{green}1} &\multicolumn{1}{c|}{\cellcolor{green}1} &\cellcolor{blue}\color{white}0 &\cellcolor{blue}\color{white}0\\
                   \hhline{~~~~~~-->{\arrayrulecolor{blue}}-->{\arrayrulecolor{black}}}
                  4 &0 &1 &1 &0 &0 &0 &0 &1 &1\\
                  5 &0 &0 &0 &0 &1 &1 &0 &0 &1\\
                  \hhline{~---------}
        \end{tabular}
        \caption{Example for upper bound calculation}
\label{ex}
\end{table}

\begin{table}
\centering
        \begin{tabular}{ c|ccccccccc|}
                  \multicolumn{1}{c}{~} & \multicolumn{1}{c}{1} & \multicolumn{1}{c}{2} & \multicolumn{1}{c}{3} & \multicolumn{1}{c}{4} & \multicolumn{1}{c}{5} & \multicolumn{1}{c}{6} & \multicolumn{1}{c}{7} & \multicolumn{1}{c}{8} & \multicolumn{1}{c}{9} \\
                   \hhline{~---------}
                  1 & \cellcolor{yellow}1  & \cellcolor{yellow}1 &\cellcolor{yellow}1 &\cellcolor{yellow}1 &\multicolumn{1}{c|}{\cellcolor{yellow}1} &0 &0 &0 &\cellcolor{yellow}0\\
                  2 &\cellcolor{yellow}1 &\cellcolor{yellow}1 &\cellcolor{yellow}1 &\cellcolor{yellow}1 &\multicolumn{0}{c|}{\cellcolor{yellow}0} &0 &0 &0 &\cellcolor{yellow}1\\
                   \hhline{~-------}
                  3 &0 &0 &0 &0 &0 &\multicolumn{1}{|c}{\cellcolor{green}1} &\multicolumn{1}{c|}{\cellcolor{green}1} &0 &0\\
                   \hhline{~~~~~~--}
                  4 &0 &\cellcolor{yellow}1 &\cellcolor{yellow}1 &0 &0 &0 &0 &\cellcolor{yellow}1 &\cellcolor{yellow}1\\
                  5 &0 &0 &0 &0 &1 &\cellcolor{green}1 &\cellcolor{green}0 &0 &\cellcolor{green}1\\
                  \hhline{~---------}
        \end{tabular}
        \caption{Solution providing the upper bound}
\label{sol}
\end{table}

\begin{enumerate}
\item
Calculate the number of ones $n_1^{in}$ and zeroes $n_0^{in}$ inside the cells of the given partial solution and the number of ones $\bar{n}_1^{out}$ and zeroes $\bar{n}_0^{out}$ which cannot get inside cells in any solution (see gray area (the gray area with black zeroes on black-and-white printing) in Table \ref{ex}).
The total number of ones $n_1$ and zeroes $n_0$ are constant.
From these values we get the numerator $a_c = n_1^{in}$ and the denominator $b_c = n_1 + n_0^{in}$ for the efficacy $f = a_c / b_c$ of the current partial solution.
For the example in Table \ref{ex} we have: $n_1^{in} = 11, n_0^{in} = 1, \bar{n}_1^{out} = 0, \bar{n}_0^{out} = 9, n_1 = 19, n_0 = 26, a_c = 11, b_c = 20$.

\item
For every unassigned machine (row) using Algorithm \ref{alg} we compare all possible alternatives of adding it to one of the existing cells or to a new cell.

For our example we have 3 alternatives for machine $i = 4$: 1) $(4, 3)$ - add it to cell 1 with parts 1, 2, 3, 4, 5, 8, 9 putting 4 ones and 3 zeroes inside this cell; 2) $(2, 2)$ - add it to cell 2 with parts 6, 7, 8, 9 putting 2 ones and 2 zeroes inside; 3) $(2, 0)$ - add it to a new cell 3 with parts 8, 9 putting 2 ones and 0 zeroes inside.
Obviously alternative 3 is better than alternative 2.
So we need to compare only two alternatives $(a_1, b_1) = (2, 0)$ and $(a_2, b_2) = (4, 3)$.
We have: $n_1^i = 4, n_0^i = 5, \Delta a = 2, \Delta b = 3, l = a_c / b_c = 11 / 20, u = (n_1 - \bar{n}_1^{out} - n_1^i) / b_c = 15/20, b_l = 20, b_u = n_1 + n_0 - \bar{n}_0^{out} - n_0^i = 31$.
And the values we need to apply Algorithm \ref{alg} are:
\[
b_1 \frac{\Delta a}{\Delta b} - a_1 = -2, \quad b_l \left( l - \frac{\Delta a}{\Delta b} \right) = -\frac{7}{3}, \quad b_u \left( u - \frac{\Delta a}{\Delta b} \right) = \frac{31}{12}
\]
So neither of the conditions in Algorithm \ref{alg} is satisfied and we cannot determine which alternative is better (Algorithm \ref{alg} returns $-1$).
In this case we build an alternative $(\max(a_1, a_2), \min(b_1, b_2))$, which is better than both incomparable alternatives, and use it to obtain an upper bound on the solution of the relaxed CFP problem.
In our example it is alternative $(4, 0)$.

Now for machine $i = 5$ we have: $n_1^i = 3, n_0^i = 6, l = 11 / 20, u = 16/20, b_l = 20, b_u = 30$.
There are 3 alternatives $(2, 4)$, $(2, 1)$, and $(1, 0)$.
It is clear that alternative $(2, 4)$ is worse than $(2, 1)$.
For $(a_1, b_1) = (1, 0)$ and $(a_2, b_2) = (2, 1)$ we have:
\[
b_1 \frac{\Delta a}{\Delta b} - a_1= -1, \quad b_l \left( l - \frac{\Delta a}{\Delta b} \right) = -9, \quad b_u \left( u - \frac{\Delta a}{\Delta b} \right) = -6
\]
So $b_u \left( u - \frac{\Delta a}{\Delta b} \right) \le b_1 \frac{\Delta a}{\Delta b} - a_1$ and Algorithm \ref{alg} returns alternative $(a_2, b_2) = (2, 1)$.
Thus $(2, 1)$ is the best alternative for machine 5.

\item
For every unassigned part (column) in the same way using Algorithm \ref{alg} we compare all possible alternatives of adding it to one of the existing cells or leaving it unassigned.
However in this case we take into account only ones and zeroes which lie in the rows already assigned in the given partial solution (blue area (the darkest area with white digits on black-and-white printing) in Table \ref{ex}).

In our example part 8 has only zeroes in this area and so it is better not to add it to any cell.
For part 9 ($j = 9$) we have 3 alternatives: 1) $(1, 1)$ - add it to cell 1 putting 1 one and 1 zero inside; 2) $(0, 1)$ - add it to cell 2 putting 0 ones and 1 zero inside; 3) $(0, 0)$ - do not add it to any cell.
It is clear that $(0, 1)$ is a bad alternative and we need to compare only $(a_1, b_1) = (0, 0)$ and $(a_2, b_2) = (1, 1)$.
We have $n_1^j = 1, n_0^j = 2, l = a_c / b_c = 11 / 20, u = (n_1 - \bar{n}_1^{out} - n_1^j) / b_c = 18 / 20, b_l = 20, b_u = n_1 + n_0 - \bar{n}_0^{out} - n_0^j = 34$.
The values needed to apply Algorithm \ref{alg} are:
\[
b_1 \frac{\Delta a}{\Delta b} - a_1 = 0, \quad b_l \left( l - \frac{\Delta a}{\Delta b} \right) = -9, \quad b_u \left( u - \frac{\Delta a}{\Delta b} \right) = -3.4
\]
So $b_u \left( u - \frac{\Delta a}{\Delta b} \right) \le b_1 \frac{\Delta a}{\Delta b} - a_1$ and Algorithm \ref{alg} returns alternative $(a_2, b_2) = (1, 1)$ as the best alternative for part 9.

\item
We calculate the upper bound by putting inside all the ones and zeroes corresponding to the best alternatives chosen for all  unassigned machines and parts.
For our example the corresponding solution which gives an upper bound to the relaxed CFP problem (and thus to the original CFP problem also) is shown in Table \ref{sol}.
For this example we have $UB = \frac{11 + 4 + 2 + 1}{19 + 0 + 1 + 1} = \frac{18}{21} \approx 0.86$.
\end{enumerate}

\section{Results}
\begin{table}
\scriptsize
\caption{Results}
\begin{tabular}{ l p{5.1cm}  p{1.0cm}  R{1.1cm}  R{1.2cm}  R{1.4cm}  R{1.4cm}  }
\hline\noalign{\smallskip}
\# & \multicolumn{1}{c}{Name} & Size & Best-known solution & \multicolumn{1}{c}{$\quad f$}  &  Time, s & Bychkov et al. (\cite{Bychkov}), Time, s \tabularnewline
\hline\noalign{\smallskip}
1 & King \& Nakornchai (\cite{6}) & $5 \times 7$ & 0.8235 & 0.8235 & 0.00 & 0.63 \tabularnewline
2 & Waghodekar \& Sahu (\cite{7}) & $5 \times 7$ & 0.6957 & 0.6957 & 0.00 & 2.29 \tabularnewline
3 & Seifoddini (\cite{8}) & $5 \times 18$ & 0.7959 & 0.7959 & 0.00 & 5.69 \tabularnewline
4 & Kusiak (\cite{9}) & $6 \times 8$ & 0.7692 & 0.7692 & 0.00 & 1.86 \tabularnewline
5 & Kusiak \& Chow (\cite{10}) & $7 \times 11$ & 0.6087 & 0.6087 & 0.00 & 9.14 \tabularnewline
6 & Boctor (\cite{11}) & $7 \times 11$ & 0.7083 & 0.7083 & 0.00 & 5.15 \tabularnewline
7 & Seifoddini \& Wolfe (\cite{12}) & $8 \times 12$ & 0.6944 & 0.6944 & 0.00 & 13.37 \tabularnewline
8 & Chandrasekharan \& Rajagopalan (\cite{13}) & $8 \times 20$ & 0.8525 & 0.8525 & 0.00 & 18.33 \tabularnewline
9 & Chandrasekharan \& Rajagopalan (\cite{14}) & $10 \times 10$ & 0.5872 & 0.5872 & 0.19 & 208.36 \tabularnewline
10 & Mosier \& Taube (\cite{15}) & $10 \times 15$ & 0.7500 & 0.7500 & 0.00 & 6.25 \tabularnewline
11 & Chan \& Milner (\cite{16}) & $10 \times 15$ & 0.9200 & 0.9200 & 0.00 & 2.93 \tabularnewline
12 & Askin \& Subramanian (\cite{17}) & $14 \times 24$ & 0.7206 & 0.7206 & 2.89 & 259.19 \tabularnewline
13 & Stanfel (\cite{18}) & $14 \times 24$ & 0.7183 & 0.7183 & 5.51 & 259.19 \tabularnewline
14 & McCormick et al. (\cite{33}) & $16 \times 24$ & 0.5326 & 0.5326 & 97117.43 & $^b$20829.38 \tabularnewline
15 & Srinivasan et al. (\cite{34}) & $16 \times 30$ & $^c$0.6899 & 0.6899 & 837.93 & $^b$13719.99 \tabularnewline
16 & King (\cite{20}) & $16 \times 43$ & 0.5753 & 0.5753 & 7045.64 & $^b$24930.93 \tabularnewline
17 & Carrie (\cite{80}) & $18 \times 24$ & 0.5773 & 0.5773 & 5668.25 & $^b$13250.01 \tabularnewline
18 & Mosier \& Taube (\cite{37}) & $20 \times 20$ & 0.4345 & $^a$0.4211 & 100000.00 & $^b$43531.77 \tabularnewline
19 & Kumar et al. (\cite{25}) & $20 \times 23$ & 0.5081 & $^a$0.4697 & 100000.00 & $^b$33020.13 \tabularnewline
20 & Carrie (\cite{80}) & $20 \times 35$ & 0.7791 & 0.7791 & 88.62 & $^b$11626.98 \tabularnewline
21 & Boe \& Cheng (\cite{50}) & $20 \times 35$ & 0.5798 & $^a$0.5615 & 100000.00 & $^b$33322.08 \tabularnewline
22 & Chandrasekharan \& Rajagopalan (\cite{130}) & $24 \times 40$ & 1.0000 & 1.0000 & 0.00 & 0.00 \tabularnewline
23 & Chandrasekharan \& Rajagopalan (\cite{130}) & $24 \times 40$ & 0.8511 & 0.8511 & 33.70 & $^b$6916.24 \tabularnewline
24 & Chandrasekharan \& Rajagopalan (\cite{130}) & $24 \times 40$ & 0.7351 & 0.7351 & 86007.93 & $^b$14408.88 \tabularnewline
25 & Chandrasekharan \& Rajagopalan (\cite{130}) & $24 \times 40$ & 0.5329 & $^a$0.5185 & 100000.00 & $^b$34524.47\tabularnewline
26 & Chandrasekharan \& Rajagopalan (\cite{130}) & $24 \times 40$ & 0.4895 & $^a$0.4648 & 100000.00 & $^b$41140.94\tabularnewline
27 & Chandrasekharan \& Rajagopalan (\cite{130}) & $24 \times 40$ & 0.4726 & $^a$0.4468 & 100000.00 & $^b$44126.76\tabularnewline
28 & McCormick et al. (\cite{33}) & $27 \times 27$ & 0.5482 & $^a$0.5017 & 100000.00 & $^b$22627.28\tabularnewline
29 & Carrie (\cite{80}) & $28 \times 46$ & 0.4706 & $^a$0.4569 & 100000.00 & $^b$71671.08\tabularnewline
30 & Kumar \& Vannelli (\cite{26}) & $30 \times 41$ & 0.6331 & $^a$0.5942 & 100000.00 & $^b$22594.20\tabularnewline
31 & Stanfel (\cite{18}) & $30 \times 50$ & 0.6012 & $^a$0.5789 & 100000.00 & $^b$31080.82\tabularnewline
32 & Stanfel (\cite{18}) & $30 \times 50$ & 0.5083 & $^a$0.4860 & 100000.00 & $^b$48977.01\tabularnewline
33 & King \& Nakornchai (\cite{6}) & $30 \times 90$ & 0.4775 & $^a$0.4684 & 100000.00 & $^b$99435.64\tabularnewline
34 & McCormick et al. (\cite{33}) & $37 \times 53$ & 0.6064 & $^a$0.5680 & 100000.00 & $^b$47744.04\tabularnewline
35 & Chandrasekharan \& Rajagopalan (\cite{120}) & $40 \times 100$ & 0.8403 & $^a$0.8403 & 100000.00 & $^b$24167.76\tabularnewline

\noalign{\smallskip}\hline
\end{tabular}
\\$^a$ The problem is not solved to optimality by our algorithm within the time limit of 100000 seconds.
\\$^b$ The problem is not solved to optimality by Bychkov et al. (\cite{Bychkov}). In this approach the problem is divided into a number of IP subproblems and a time limit of 300 seconds is set for every subproblem.
\\$^c$ A greater value is reported in some papers on heuristics probably due to an incorrect input matrix.
\label{res}
\end{table}

The suggested branch and bound algorithm has been able to solve 21 of 35 popular benchmark instances from the literature exactly and to find good solutions for the remaining 14 instances.
The results are presented in Table \ref{res}.  All computations were run on Intel Core i7 with 16Gb RAM.
Note that the algorithm was run without any initial solution while in the results reported by Bychkov et al. (\cite{Bychkov}) the best-known solutions were used as initial.
The results show that the developed algorithm is more efficient than the approach suggested by Bychkov et al. (\cite{Bychkov}).

\section{Acknowledgment}
This research is supported by Laboratory of Algorithms and Technologies for Network Analysis, NRU HSE.

\end{document}